\documentclass[sigconf]{acmart}
\usepackage{amsmath,amssymb,amsfonts}
\usepackage{algorithmic}
\usepackage{graphicx}
\usepackage{scalerel}
\usepackage{textcomp}
\usepackage{xcolor}
\usepackage{tcolorbox}
\usepackage{booktabs}
\usepackage{tikz}

\copyrightyear{2019}
\acmYear{2019}
\acmConference[SBES 2019]{XXXIII Brazilian Symposium on Software Engineering}{September 23--27, 2019}{Salvador, Brazil}
\acmBooktitle{XXXIII Brazilian Symposium on Software Engineering (SBES 2019), September 23--27, 2019, Salvador, Brazil}
\acmPrice{15.00}
\acmDOI{10.1145/3350768.3350788}
\acmISBN{978-1-4503-7651-8/19/09}

\def\BibTeX{{\rm B\kern-.05em{\sc i\kern-.025em b}\kern-.08em
    T\kern-.1667em\lower.7ex\hbox{E}\kern-.125emX}}

\newcommand{\mybox}[2]{
\begin{tcolorbox}[halign=justify, rounded corners,top=5pt,bottom=5pt,left=5pt,right=5pt,colback=white!50]
\textit{#1: }#2
\end{tcolorbox}
}

\newcommand{\aspas}[1]{{``#1''}}

\newcommand{\sbarf}[2]{{\color{darkgray}\rule{\dimexpr 1cm * #1 / #2}{6pt}\color{lightgray}\rule{\dimexpr 1cm * (#2 - #1) / #2}{6pt}}}

\def\thumbsup{\scalerel*{\includegraphics{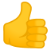}}{O}}
\def\thumbsdown{\scalerel*{\includegraphics{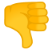}}{O}}
\def\laugh{\scalerel*{\includegraphics{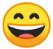}}{O}}
\def\hooray{\scalerel*{\includegraphics{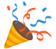}}{O}}
\def\confused{\scalerel*{\includegraphics{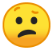}}{O}}
\def\heart{\scalerel*{\includegraphics{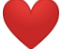}}{O}}
\def\eyes{\scalerel*{\includegraphics{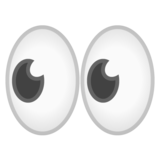}}{O}}
\def\rocket{\scalerel*{\includegraphics{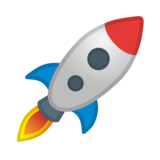}}{O}}

\def\clap{\scalerel*{\includegraphics{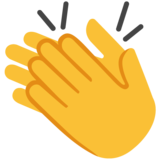}}{O}}

\begin{document}

\title{Beyond Textual Issues: Understanding the Usage and Impact of GitHub Reactions}


\author{Hudson Borges}
\affiliation{%
    \department{Faculty of Computer Science (FACOM)}
    \institution{Federal University of Mato Grosso do Sul (UFMS)}
    \city{Campo Grande}
    \state{Mato Grosso do Sul}
    \country{Brazil}
}
\email{hsborges@facom.ufms.br}

\author{Rodrigo Brito and Marco Tulio Valente}
\affiliation{%
    \department{Department of Computer Science (DCC)}
    \institution{Federal University of Minas Gerais (UFMG)}
    \city{Belo Horizonte}
    \state{Minas Gerais}
    \country{Brazil}
}
\email{{ britorodrigo, mtov } @ dcc.ufmg.br}


\begin{abstract}
Recently, GitHub introduced a new social feature, named \textit{reactions}, which are \aspas{pictorial characters} similar to \textit{emoji} symbols widely used nowadays in text-based communications. Particularly, GitHub users can use a pre-defined set of such symbols to react to issues and pull requests.
However, little is known about the real usage and impact of GitHub reactions.
In this paper, we analyze the reactions 
provided by developers to more than 2.5 million issues and 9.7 million issue comments, in order to answer an extensive list of nine research questions about the usage and adoption of reactions.
We show that reactions are being increasingly used by open source developers. Moreover, we also found that issues with reactions usually take more time to be handled and have longer discussions.
\end{abstract}


\maketitle

\section{Introduction}

Over the last years, open source software development has become more social and collaborative.
In fact, social coding fosters formal and informal collaboration in software development by empowering knowledge exchange between developers~\cite{DabbishSTH12}.
In this context, traditional code management tools (e.g., control version systems, issue tracking, etc.) have been extended with social networks features to connect developers and increase their social interactions.
Among the social coding platforms, GitHub is nowadays the most popular one and the world's largest collection of open source software, with around 36 million users and 104 million repositories.\footnote{\url{https://github.com/search}, verified on 01/20/2019.}

Recently, GitHub introduced a new social feature, named \textit{reactions}, which are \aspas{pictorial characters} similar to \textit{emoji} symbols widely used nowadays in text-based communications. 
Indeed, emojis are increasingly being used in many areas and are a popular mechanism for developers express their emotions.
However, little is known about the real usage and impact of GitHub reactions on software development practices.

In this paper, we investigate how developers are using the reactions feature introduced by GitHub. To this purpose, we collect and analyze reactions to more than 2.5 million GitHub issues and 9.7 million comments associated to these issues.
In particular, we ask nine research questions, which are grouped into two major themes:

\smallskip\noindent{Theme 1: Usage of reactions}
\newline\noindent\textit{RQ \#1: Do developers use reactions?}
\newline\noindent\textit{RQ \#2: What are the most common reactions?}
\newline\noindent\textit{RQ \#3: How is the usage of reactions evolving?}
\newline\noindent\textit{RQ \#4: Do popular projects have more reactions?}
\newline\noindent\textit{RQ \#5: Do certain types of issues have more reactions?}

\smallskip\noindent{Theme 2: Impact of Reactions}
\newline\noindent\textit{RQ \#6: Do issues with more reactions get resolved faster?}
\newline\noindent\textit{RQ \#7: Do issues with more reactions have more discussion?}
\newline\noindent\textit{RQ \#8: Do negative reactions inhibit further participation?}
\newline\noindent\textit{RQ \#9: Do reactions reduce the noise in issue discussions?}\smallskip

We show that more than 28\% of the issues threads (i.e., an issue and its respective comments) receive at least one reaction. Moreover, the usage of reactions is growing fast in the last two years. \textit{Thumbs up} is the most common reaction, with almost 80\% of all reactions. By contrast, \textit{Confused} is the least common one, with just 1.7\%. Also, we show that popularity does not have a major influence on the reactions whereas programming language and issue type do have.

Regarding the impact on software development, we found that issues with reactions take more time to be handled and have longer discussions. Moreover, negative reactions do not inhibit the participation of developers in further discussions. Finally, we show that reactions are contributing to reduce trivial comments (i.e., those containing only an emoji).

To our knowledge, we are the first to study GitHub reactions. Particularly, our contributions are twofold: (1) a thorough quantitative analysis on the usage and impact of GitHub reactions, focusing on their usage in issues; 
(2) a list of practical recommendations for using reactions, both for project owners and users. With these recommendations, we hope to demystify the use of reactions. On the one hand, we provide empirical data to convince developers to use reactions in recommended scenarios; on the other hand, we prevent the set up of unrealistic expectations regarding the benefits of reactions. Finally, we are making public the dataset used in this paper, with reactions applied to more than 2.5 million issues and 9.7 million comments.

The remainder of this paper is organized as follows.
In Section~\ref{sec:background}, we present background information about GitHub reactions. Next, in Section~\ref{sec:study-design}, we describe the design of the study presented in this paper. Then, in Section~\ref{sec:results} we answer the proposed research questions. Our key findings are described in Section~\ref{sec:key-finding-and-implications} and threats to validity are discussed in Section~\ref{sec:threats-to-validity}. Finally, we present related work in Section~\ref{sec:related} and we conclude the paper in Section~\ref{sec:conclusion}.
\section{Background}
\label{sec:background}

GitHub implements a vast number of social features. For example, developers can follow other developers to receive notifications about events on the platform. Moreover, developers can fork projects to create a copy they can freely modify without affecting the original code \cite{Jiang2017}. Finally, GitHub provides a social feature named \textit{stars} \cite{borges2016}, which is similar to a \textit{like} in social media sites (e.g., Facebook, Twitter, HackerNews, Reddit, etc.).

\begin{figure}[!ht]
    \centering
    \includegraphics[width=\linewidth, trim={0 0 0 0em}, clip]{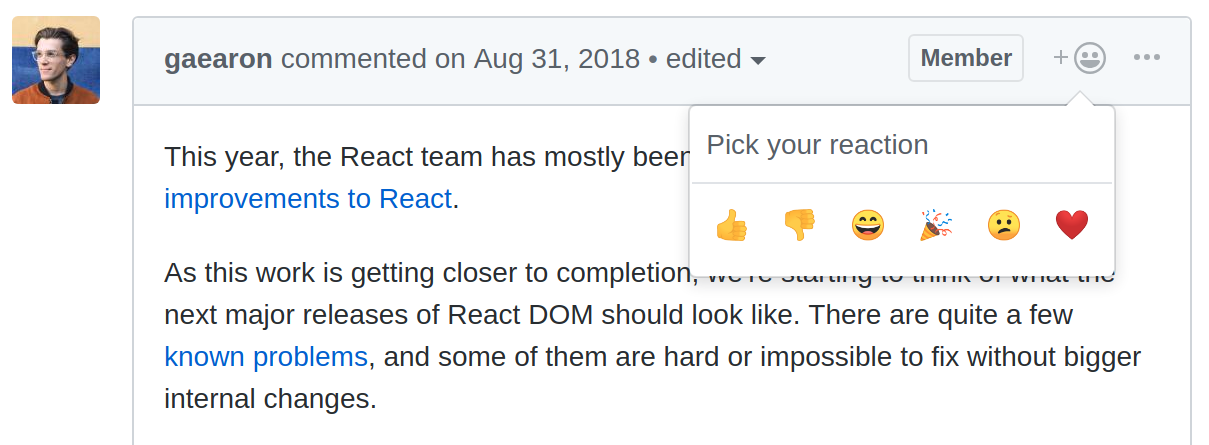}
    \caption{Reacting to a comment on GitHub}
    \label{fig:github-reactions} 
\end{figure}

In 2016, GitHub introduced a new social feature named \textit{Reactions} to allow developers quickly express their feelings on issues and pull requests (Figure~\ref{fig:github-reactions}).
GitHub designers argued there are complex and detailed comments that require textual messages, but sometimes developers can use a simple \textit{emoji} to express their feelings on someone's comment.
Besides that, they argue that reacting to a comment avoids long threads with many comments lacking relevant content.

GitHub provides six different reactions for use in issues and pull requests, which are showed in Table \ref{tab:reactions} and described next.\footnote{\url{https://developer.github.com/v3/reactions}, verified on 01/20/2019.}

\medskip\noindent\textit{Thumbs up}, which is used to express acceptance or approval.

\smallskip\noindent\textit{Thumbs down}, which is used to express rejection or disapproval.

\smallskip\noindent\textit{Laugh}, which indicates a fun situation or happiness. It is also used to express sarcasm or irony in negative situations.

\smallskip\noindent\textit{Hooray}, which is also used to celebrate or express a happy situation. For example, it is commonly used in software releases.

\smallskip\noindent\textit{Confused}, which is proposed to express doubt or some difficulty to understand a specific subject.

\smallskip\noindent\textit{Heart}, which indicates a strong acceptance or approval; it is used when the user really likes something.\smallskip

\begin{table}[!ht]
\centering
\caption{GitHub reactions}
\label{tab:reactions}
\resizebox{0.7\columnwidth}{!}{
\begin{tabular}{@{}llll@{}}
\toprule
Name        & Symbol & Name & Symbol \\ \midrule
Thumbs up   & \includegraphics[width=15pt]{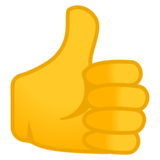} & Hooray      & \includegraphics[width=15pt]{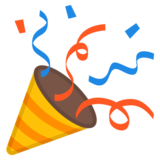} \\
Thumbs down & \includegraphics[width=15pt]{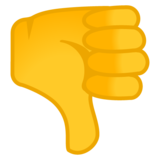} & Confused    & \includegraphics[width=15pt]{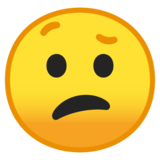} \\
Laugh       & \includegraphics[width=15pt]{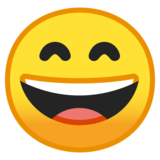} & Heart        & \includegraphics[width=15pt]{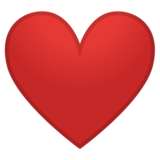} \\ \bottomrule
\end{tabular}}
\end{table}

Developers can add only one reaction per textual element.
Also, reactions can be added to closed issues and pull requests.
However, project owners can lock issues or pull requests to disable both comments and reactions.
\section{Study Design}
\label{sec:study-design}

In this study, we restrict our analysis to reactions added to issues and their respective comments. 
Although pull requests can also receive reactions from developers, they typically have different purposes and characteristics, which should be better explored in further work.
In the next paragraphs, we describe the criteria to select the projects, issues, and reactions studied in this paper. We also characterize our final dataset.

\subsection{Initial projects sampling}

To study how developers apply reactions to GitHub issues, we first defined an initial sample of projects.
This sample includes the top-5,000 repositories with more stars on GitHub (stars are considered a reliable proxy for the popularity of GitHub projects~\cite{borges2016, borges2018}).
To obtain the metadata about these projects, we used GitHub GraphQL API.\footnote{\url{https://developer.github.com/v4}, verified on 01/20/2019} Here, we used the \textit{search} function to navigate through the first 5,000 repositories by stars. For each repository, we retrieved basic metadata, such as programming language, creation date, last update date, and owner.

\subsection{Issues collection}

After collecting the initial sample of projects, we again used GitHub's GraphQL API to obtain the issues created in these projects after March 13rd, 2016.
We restricted the issues collection to this date because it was when the reaction feature was officially announced.\footnote{\url{https://blog.github.com/2016-03-10-add-reactions-to-pull-requests-issues-and-comments}, verified on 01/20/2019.}
For each issue, we collected the date when it was created, updated, and closed. In addition, we also collected data about the user who created the issue and his/her role in the project (e.g., collaborator, contributor, owner, or none). Finally, we retrieved all the events in the issue timeline. For example, an event can be a comment, a reference to a commit or another issue, and the action of closing or reopening the issue.
Finally, we also collected data about the reactions of each issue and their respective comments. This data includes: (i) the user who reacted; (ii) the reaction (as detailed in Section~\ref{sec:background}); and (iii) the reaction date.

After the data collection, we found that some projects do not have issues created after March 13rd, 2016.
Specifically, we found that 159 out of 5,000 projects (3.18\%) have no issues after this date. Probably, they are deprecated or feature-completed projects.
Thus, we excluded these projects from this study.
In total, we collected 2,544,304 issues and 9,775,928 comments, from 4,841 GitHub projects.

\subsection{Final dataset characterization}

\begin{figure}[!ht]
    \centering
    \includegraphics[width=.95\linewidth, trim={0 0 0 2em}, clip]{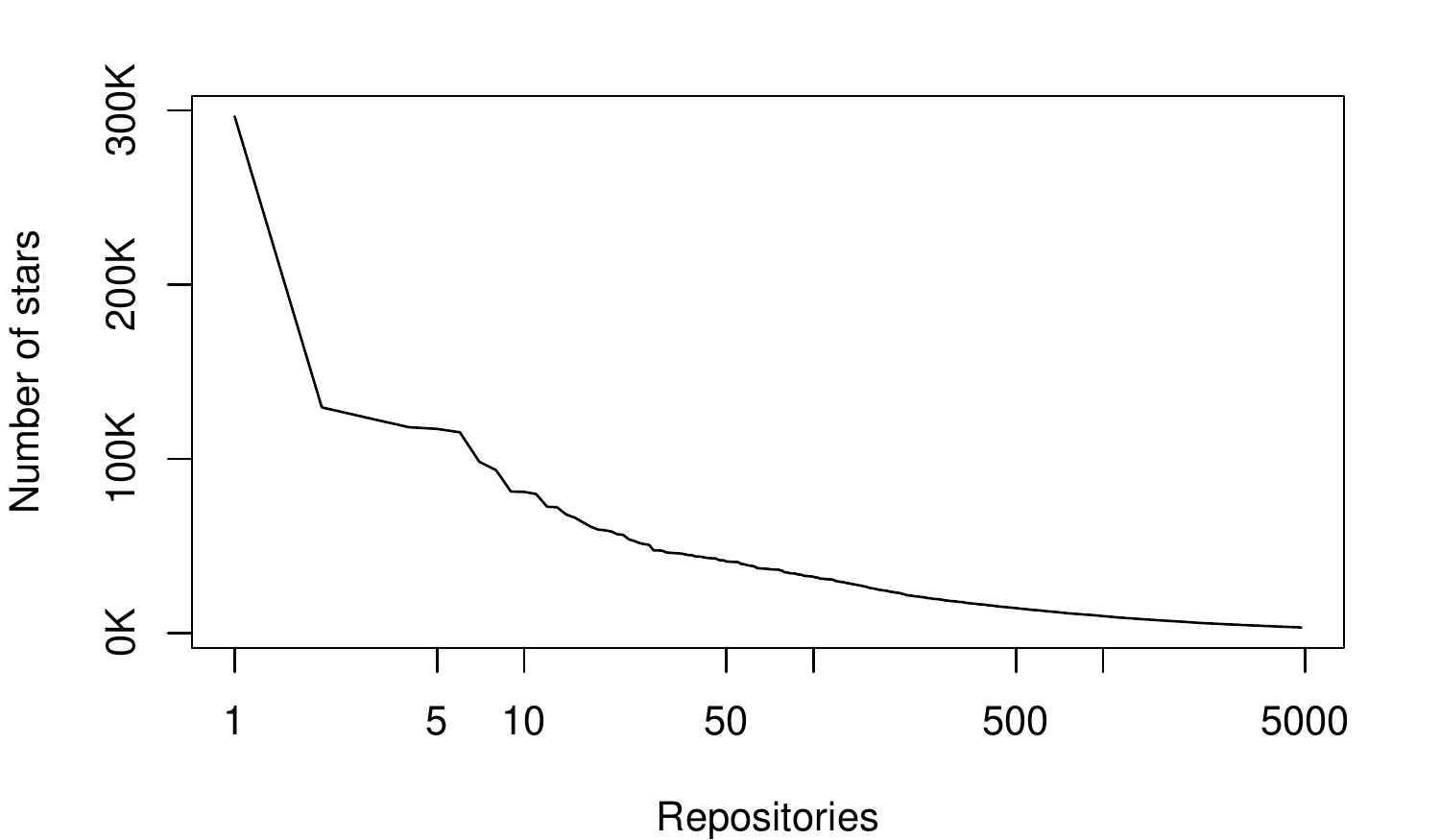}
    \caption{Popularity of the analyzed repositories}
    \label{fig:design:popularity}
\end{figure}

Our final dataset consists of 4,841 public GitHub projects.
Regarding their popularity, the number of stars ranges from 296K stars (\textsc{freeCodeCamp/freeCodeCamp}) to 3,038 stars (\textsc{sindresorhus/\\speed-test}). Figure~\ref{fig:design:popularity} shows the distribution of the number of stars of these projects. Other popular projects in our dataset include \textsc{twbs/bootstrap} (129K stars), \textsc{vuejs/vue} (123K stars), and \textsc{facebook/react} (118K stars). In summary, our study is based on relevant and widely used open source projects.

\begin{figure}[!ht]
    \centering
    \includegraphics[width=.95\linewidth, trim={0 0.5em 0 2em}, clip]{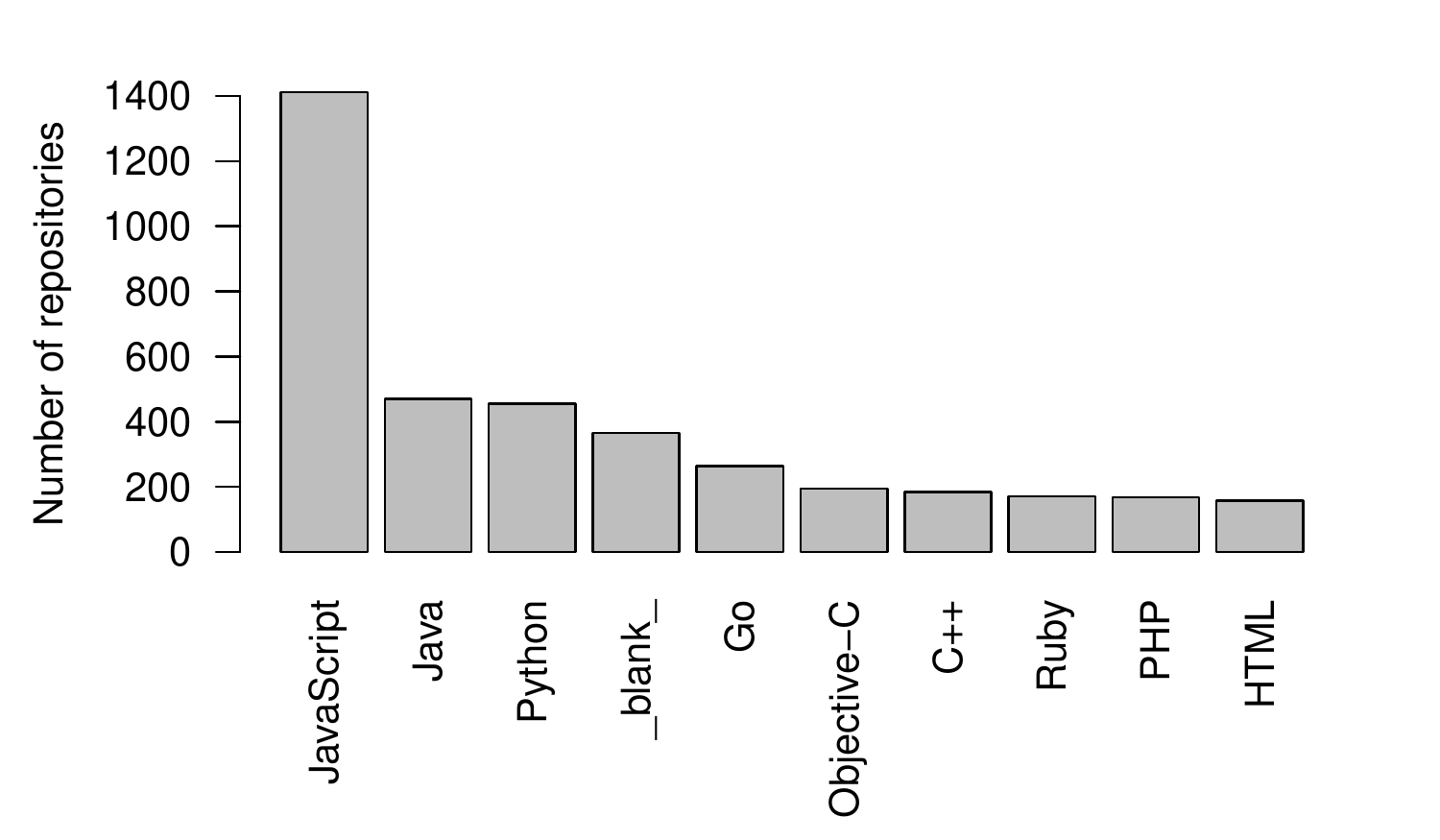}
    \caption{Top-10 programming language with more repositories}
    \label{fig:design:languages}
\end{figure}

Considering the primary programming language of these projects, JavaScript has the highest number of repositories (1,431 repositories), followed by Java (470 repositories) and Python (456 repositories); see details in Figure~\ref{fig:design:languages}. The prevalence of JavaScript repositories is common on studies based on GitHub projects~\cite{borges2016}. Moreover, 366 projects have no programming language, i.e., they are documentation projects, such as books, tutorials, awesome-lists, etc. In total, the projects in our dataset are implemented in 70 different programming languages. Therefore, although JavaScript dominates, we also have a variety of languages, which allows us to study reactions in several of ecosystems.

\begin{figure}[!ht]
    \centering
    \includegraphics[width=.8\linewidth, trim={3em 0 1em 2em}, clip]{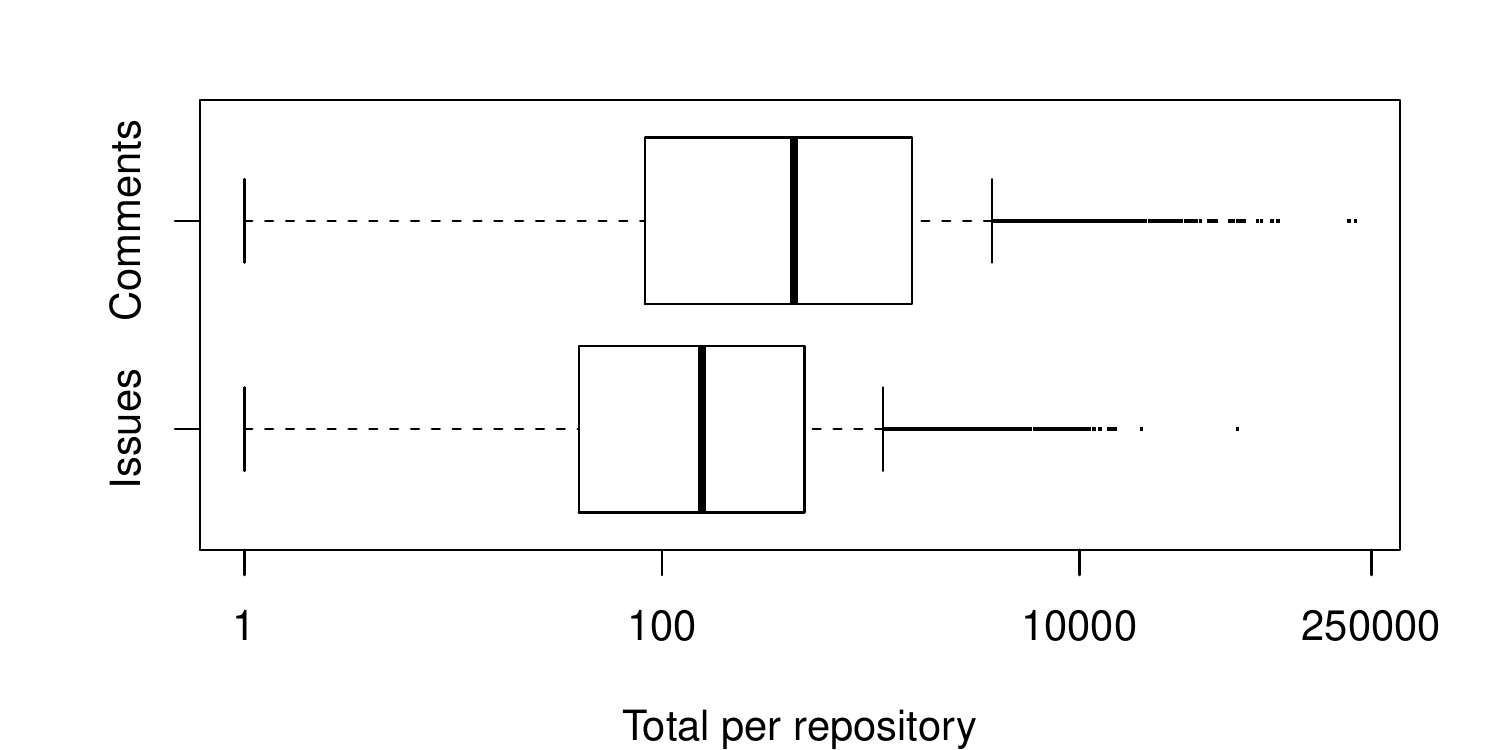}
    \caption{Number of issues and comments per repository}
    \label{fig:design:issues_comments}
\end{figure}

In GitHub, issues have a textual description, which report a bug, feature, enhancement, etc. Project members then discuss the reported issue, by means of comments. Figure~\ref{fig:design:issues_comments} shows the distribution of the number of issues and comments in our dataset. The repository with the highest number of issues is \textsc{Microsoft/vscode} (57K issues), followed by \textsc{kubernetes/kubernetes} (19.8K issues) and \textsc{Microsoft/TypeScript} (14.9K issues). Regarding the number of comments, \textsc{Microsoft/vscode} also has the highest result (210K comments), followed by \textsc{kubernetes/kubernetes} (196K comments) and \textsc{golang/go} (89.4K comments).
On average, each issue has 3.84 comments.

\section{Results}
\label{sec:results}

\vspace{1em}\begin{tcolorbox}[halign=flush center, sharp corners,top=1pt,bottom=1pt]
Theme 1: Usage of reactions
\end{tcolorbox}

\vspace{0.5em}\noindent\textit{RQ \#1: Do developers use reactions?}\vspace{0.5em}

\begin{figure}[!ht]
    \centering
    \includegraphics[width=.8\linewidth, trim={3em 0 1em 2em}, clip, page=1]{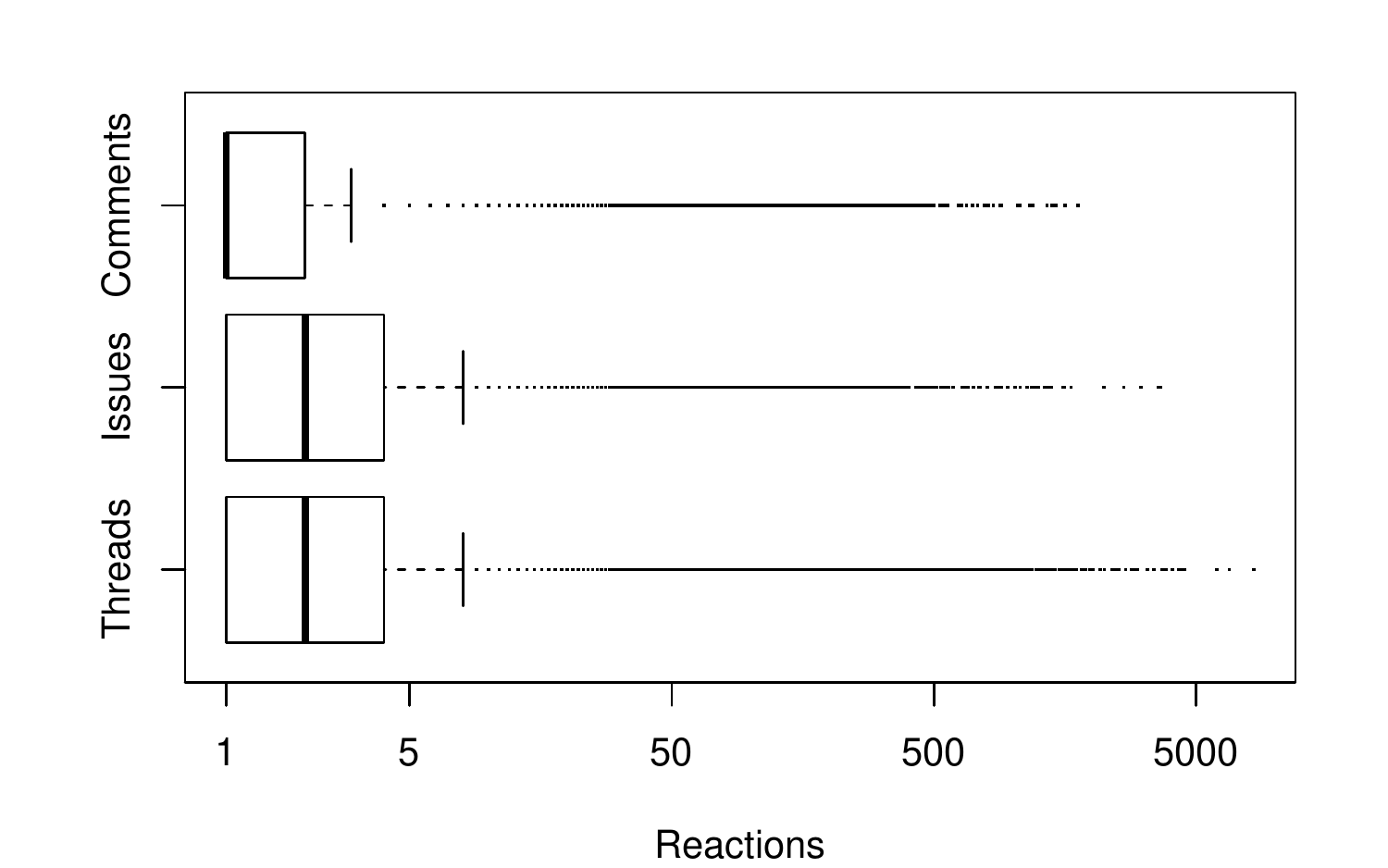}
    \caption{Number of reactions per issue, comment, and thread}
    \label{fig:results:usage}
\end{figure}

In this initial research question, our goal is to quantify the usage of reactions in GitHub issues. To this purpose, we first compute the number of reactions in our dataset.
We found that most GitHub issues (87.0\%) and comments (89.3\%) do not receive reactions.
However, when considering whole issues threads, i.e., the initial document that describes an issue and the associated comments, we found that 28.4\% have at least one reaction.
Figure~\ref{fig:results:usage} shows the number of reactions in issues (i.e., on the initial document that describes an issue) and comments (i.e., on further messages that discuss the issue), and in whole issue threads. 
Furthermore, in the box plot we only consider elements with at least one reaction.
The 1st, median, and 3rd quartiles for reactions/issue and reactions/thread are 1, 2, and 4. Considering reactions/comments, the same measures are 1, 1, and 2.

\begin{figure}[!ht]
 \centering
 \includegraphics[width=\linewidth, clip]{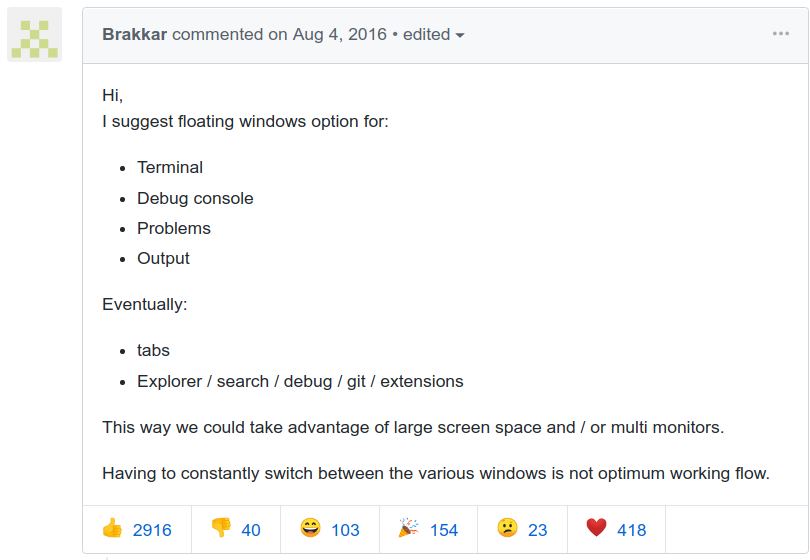}
 \caption{Issue \aspas{\#10121 -- Allow for floating windows} in \textsc{Microsoft/vscode}. Reactions appear in the bottom.}
 \label{fig:most-reacted-issue}
\end{figure}

Figure~\ref{fig:most-reacted-issue} shows the most reacted issue in our dataset. This issue, titled \aspas{\it Allow for floating windows}, belongs to \textsc{Microsoft/vscode}, has 3,654 reactions, and it is opened since August 4, 2016. It is a request to add floating windows in the \textit{Visual Studio Code} editor. Despite the high number of reactions, the issue has not been fully implemented by the project maintainers. In terms of comments, the most reacted one is issue \aspas{\it \#980 -- WebAssembly logo voting}, from \textsc{WebAssembly/design}. This issue was created with the goal of selecting a new logo for the project. The candidate logos were presented in the comments and the most reacted one was chosen. This "winner" comment received 1,770 reactions (1,486 \textit{Thumbs up}, 120 \textit{Thumbs down}, 11 \textit{Laugh}, 46 \textit{Hooray}, 34 \textit{Confused}, and 73 \textit{Heart}).

\begin{figure}[!ht]
    \centering
    \includegraphics[width=\linewidth, trim={0 0.75em 0 2em}, clip, page=1]{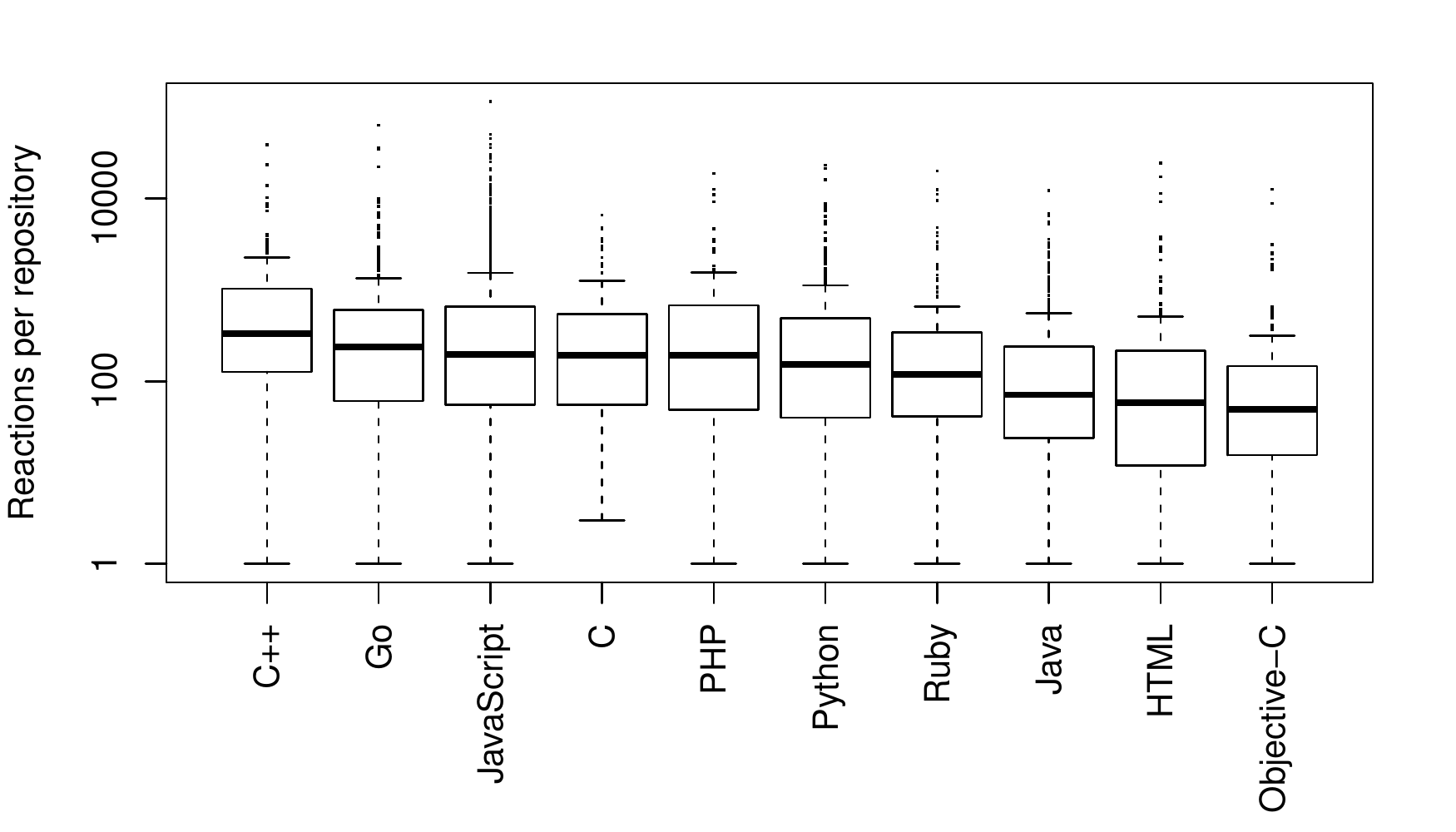}
    \caption{Number of reactions per repository}
    \label{fig:results:usage-language}
\end{figure}

Finally, we analyzed the usage of reactions by programming language. Here, we computed the total number of reactions applied to a repository (i.e., counting reactions added to issues and comments) and compared the distributions, by programming language.
Figure~\ref{fig:results:usage-language} presents the distributions of the top-10 languages with more reactions by repository. We observe that projects implemented in C++ receive more reactions than projects in other languages. By contrast, Objective-C projects are the ones with the lowest number of reactions.
By applying the Kruskal-Wallis test to compare multiple samples, we found that these distributions differ in at least one language (\emph{p-value} $<$ 0.001).
Then, we use a non-parametric, pairwise and multiple comparisons test (Dunn's test) to isolate the languages that differ from the others.
Indeed, we found that the distribution of C++ projects is different from all other distributions. Additionally, the distribution of Objective-C projects differ from all others, except HTML. 
Our hypothesis is that Objective-C projects have less reactions because iOS development is moving to Swift. Therefore, Objective-C projects are gradually becoming deprecated or obsolete.

\vspace{0.5em}\mybox{Summary}{Most issues (87\%) and comments (89\%) do not have reactions. By contrast, there are issues with 3,654 reactions. For whole threads (issues + comments), we found that 28.4\% have at least one reaction.}

\vspace{0.5em}\noindent\textit{RQ \#2: What are the most common reactions?}\vspace{0.5em}

This second research question aims to reveal the most common reactions and their roles in GitHub issues.
Table~\ref{tab:most-used} shows the results. \textit{Thumbs up} is the most common reaction (78.7\%), followed by \textit{Hooray} and \textit{Heart}, both with just 5.9\%. \textit{Confused} is the least common reaction, with just 1.7\%. Table~\ref{tab:most-used} also presents information about the issue with the highest number of reactions. 
For \textit{Thumbs up}, the highest value is found in the issue described in RQ \#1 (Figure~\ref{fig:most-reacted-issue}). 
For \textit{Hooray} and \textit{Heart}, the highest values are 1,031 and 513 reactions, respectively, which are found in the issue \aspas{\it React 16 RC}\footnote{https://github.com/facebook/react/issues/10294} from \textsc{facebook/react}. The reasons for the high acceptance of this issue includes a complete rework of the framework's API and the addition of long-standing feature requests.
By contrast, the issue \aspas{\it Changing Redis master-slave replication terms with something else},\footnote{https://github.com/antirez/redis/issues/5335} which was created by the repository owner, received the highest number of \textit{Thumbs down} and \textit{Confused} reactions.
According to the comments in this issue, the proposal does not add value to the project and causes compatibility issues with previous versions.
Finally, issue \aspas{\it Check continuity on O2 cryogenic tanks before allowing stir}\footnote{https://github.com/chrislgarry/Apollo-11/issues/3} from \textsc{chrislgarry/Apollo-11} received the highest number of \textit{Laugh} reactions.
Clearly, this issue was only created to cause laughter. The repository in question archives the original source code of Apollo-11's command and lunar modules.

\begin{table}[!ht]
\centering
\caption{Most common GitHub reactions}
\label{tab:most-used}
\resizebox{0.478\textwidth}{!}{
\begin{tabular}{@{}lrlrr@{}}
\toprule
\multicolumn{1}{c}{Reaction} & \multicolumn{1}{c}{Ratio (\%)} & \multicolumn{1}{c}{Repository} & \multicolumn{1}{r}{Issue} & \multicolumn{1}{c}{Max} \\
\midrule
Thumbs up    & 78.7 \sbarf{7873}{10000} & Microsoft/vscode & 10121  & 2,916 \\
Hooray       & 5.9  \sbarf{593}{10000}  & facebook/react & 10294  & 1,031 \\
Heart        & 5.9  \sbarf{586}{10000}  & facebook/react & 10294  & 513 \\
Thumbs down  & 5.0  \sbarf{505}{10000}  & antirez/redis   & 5335   & 683 \\
Laugh        & 2.7  \sbarf{274}{10000}  & chrislgarry/Apollo-11 & 3 & 1,083 \\
Confused     & 1.7  \sbarf{169}{10000}  & antirez/redis & 5335 & 155 \\
\bottomrule
\end{tabular}
}
\end{table}

\vspace{0.5em}\mybox{Summary}{Thumbs Up is by far the most common reaction (78.7\%), followed by Hooray (5.9\%), Heart (5.9\%), and Thumbs down (5.0\%).}

\vspace{1em}\noindent\textit{RQ \#3: How is the usage of reactions evolving?}\vspace{0.5em}

\begin{figure}[!ht]
    \centering
    \includegraphics[width=.9\linewidth, trim={0 2 0 2.5em}, clip,page=1]{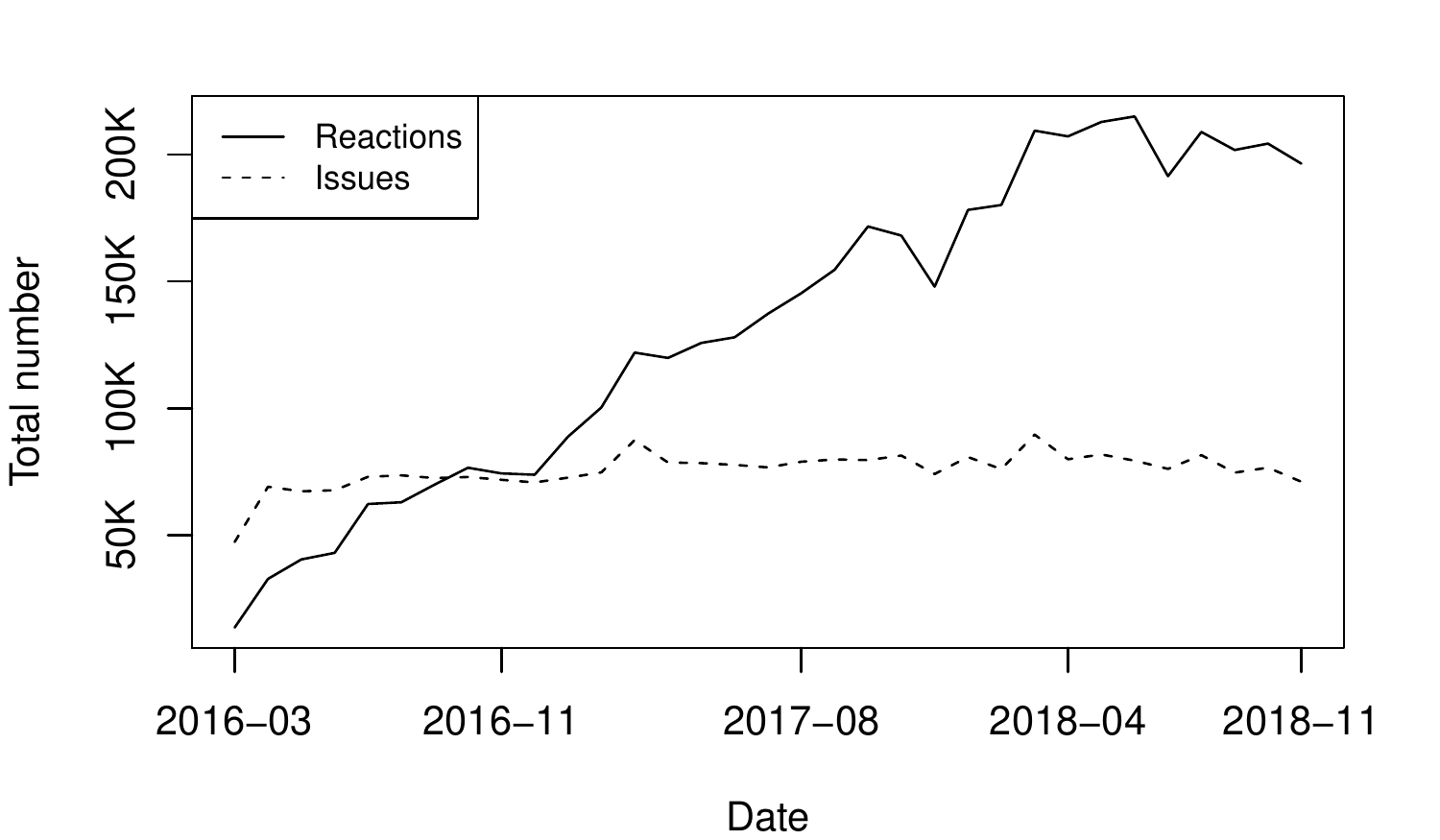}
    \caption{New reactions and issues per month}
    \label{fig:results:evolution}
\end{figure}

In this third research question, we investigate the evolution of the usage of GitHub reactions.
Specifically, we compared the number of issues opened with the number of reactions since this feature was introduced in GitHub.
Figure~\ref{fig:results:evolution} shows the evolution of both measures.
The curves reveal that reactions usage is growing fast since its introduction, although with a stabilization trend in the last months. By contrast, the number of new issues per month remains almost constant since 2016.

\begin{figure}[!ht]
    \centering
    \includegraphics[width=.9\linewidth, trim={0 0 0 2.5em}, clip,page=2]{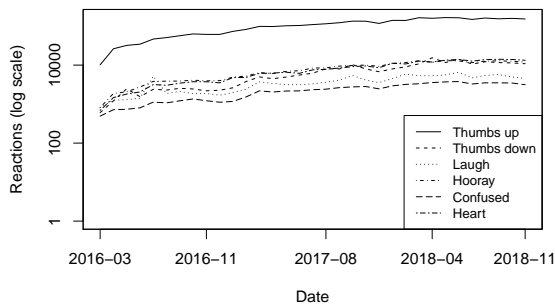}
    \caption{New reactions per month (per reactions type)}
    \label{fig:results:evolution2}
\end{figure}

Finally, Figure~\ref{fig:results:evolution2} shows the growth of each reaction type.
As we can see, there is no major difference on the growth pattern of existing reactions. The exception is the \textit{Laugh} reaction, which started as the second most used reaction in 2016 (8.3\%) and is now the second least used one (2.3\%).

\vspace{0.5em}\mybox{Summary}{Reactions are gaining momentum; in 2018, developers used 32.5\% more reactions than in 2017.}

\vspace{0.5em}\noindent\textit{RQ \#4: Do popular projects have more reactions?}\vspace{0.5em}

\begin{figure}[!ht]
    \centering
    \includegraphics[width=0.75\linewidth, trim={0 0 0 2.5em}, clip]{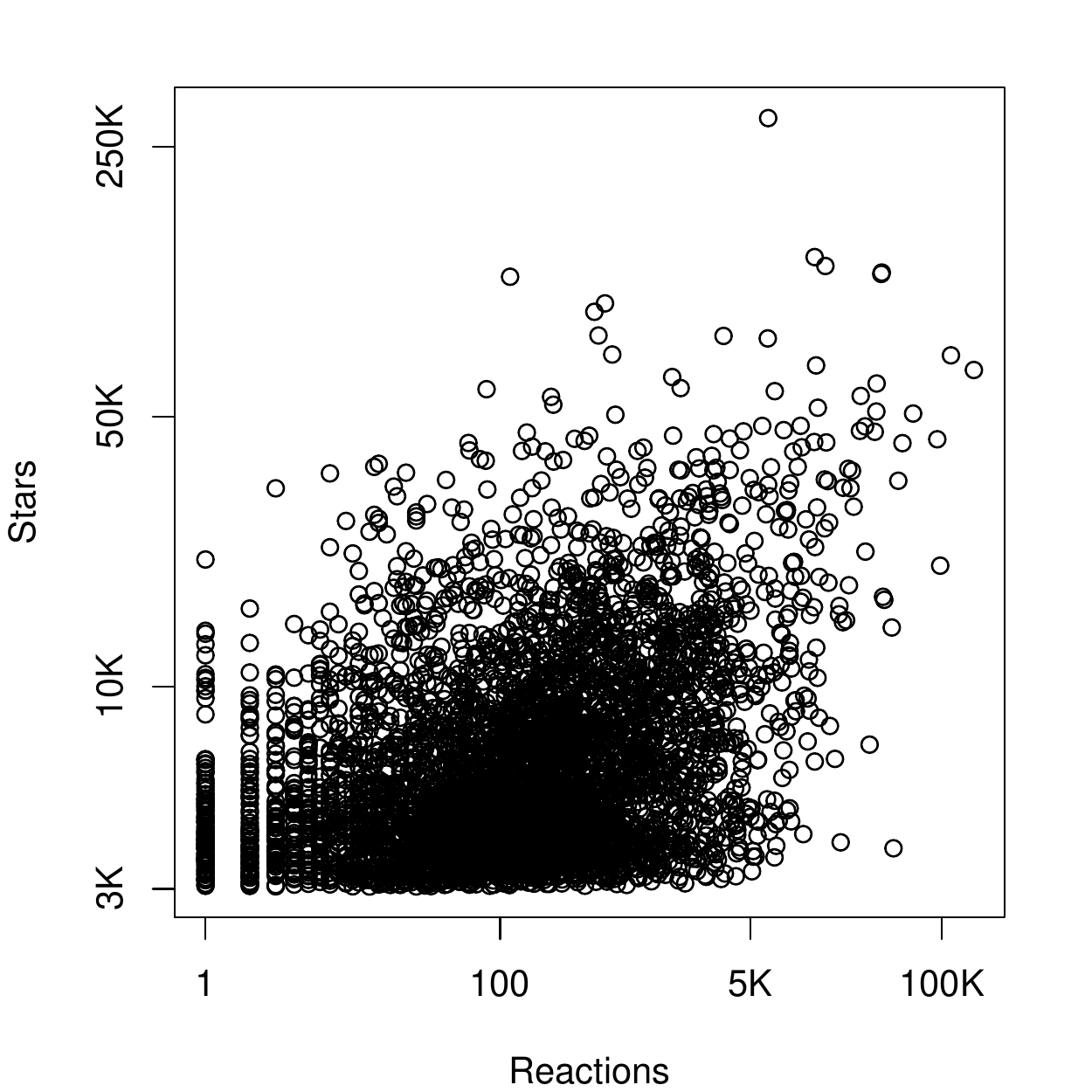}
    \caption{Stars vs reactions correlation}
    \label{fig:results:usage-top}
\end{figure}

In this research question, we examine whether issues from popular projects have more reactions.
Figure~\ref{fig:results:usage-top} shows a scatterplot correlating the total number of reactions with the number of stars of the 4,841 repositories in our dataset.
The result suggests that there is no major correlation between these metrics.
We ran Spearman's rank correlation test and the resulting correlation coefficient $\mathit{rho}$ is 0.340, with $\emph{p-value} <$ 0.001.
This $rho$ value denotes a low correlation.

\vspace{0.5em}\mybox{Summary}{The projects popularity (measured by GitHub stars) are weakly correlated to the number of reactions received.}

\vspace{0.5em}\noindent\textit{RQ \#5: Do certain types of issues have more reactions?}\vspace{0.5em}

In this research question, we analyze how the use of reactions varies according to the type of issues (e.g., bugs, enhancements, etc.).
To this purpose, we rely on the labels provided by developers when reporting issues on GitHub. 
In total, we found 22,856 distinct labels in 1,252,926 issues (i.e., 49.2\% of the issues in our dataset). Table~\ref{tab:tags-list} lists the top-20 most common issues labels. As we can see, the top-3 are \textit{bug} (14.4\%), \textit{question} (6.8\%), and \textit{enhancement} (6.4\%).

\begin{table}[!ht]
\centering
\caption{Top-20 most used labels on GitHub issues}
\label{tab:tags-list}
\resizebox{0.478\textwidth}{!}{
\begin{tabular}{@{}lr|lr@{}}
\toprule
\multicolumn{1}{c}{Label} & \multicolumn{1}{c}{Issues} & \multicolumn{1}{c}{Label} & \multicolumn{1}{c}{Issues} \\
\midrule
bug             & 179,840 \sbarf{179840}{1252926} & 
feature request & 16,307 \sbarf{16307}{1252926}  \\
question    & 84,589  \sbarf{84589}{1252926} &
good first issue & 15,524 \sbarf{15524}{1252926}  \\
enhancement & 80,270  \sbarf{80270}{1252926}  &
documentation & 13,170 \sbarf{13170}{1252926}  \\
help wanted & 42,619  \sbarf{42619}{1252926}  &
wontfix & 12,577 \sbarf{12577}{1252926}  \\
duplicate   & 28,850  \sbarf{28850}{1252926}  &
feature-request & 12,333 \sbarf{12333}{1252926}  \\
stale       & 23,788  \sbarf{23788}{1252926}  &
needs more info & 10,894 \sbarf{10894}{1252926}  \\
kind/bug    & 21,357  \sbarf{21357}{1252926}  &
support & 10,819 \sbarf{10819}{1252926}  \\
type: bug   & 21,119  \sbarf{21119}{1252926}  &
verified & 10,174 \sbarf{10174}{1252926}  \\
invalid     & 19,673  \sbarf{19673}{1252926}  &
discussion & 9,663 \sbarf{9663}{1252926}  \\
feature     & 19,120  \sbarf{19120}{1252926}  &
*duplicate & 9,024 \sbarf{9024}{1252926}  \\
\bottomrule
\end{tabular}
}
\end{table}

However, since GitHub allows developers to create their own labels, it is common to have labels sharing the same purpose. For example, \textit{bug}, \textit{kind/bug}, and \textit{type: bug} all refer to bugs.
Thus, we performed an extra manual step to group labels with the same purpose.
After this step, we restrict the analysis to eight main labels: \textit{bug}, \textit{duplicate}, \textit{enhancement}, \textit{invalid}, \textit{question}, \textit{wontfix}, \textit{help}, and \textit{stale}.
This selection covers 68.6\% of the issues with labels initially considered in this RQ.

Table~\ref{tab:tags-summary} shows the number of issues that have these labels and the total number of reactions received by them. As mentioned before, \textit{bug} is the most common type of issue (column \textit{Issues}). However, the number of reactions per \textit{bug} (0.35) is almost four time lower than the number of reactions per \textit{enhancements} (1.39).
Interestingly, this finding suggests that GitHub users tend to react more to enhancements than to bugs.

\begin{table}[!ht]
\centering
\caption{Issues labels}
\label{tab:tags-summary}
\begin{tabular}{@{}lrrr@{}}
\toprule
\multicolumn{1}{c}{Label} & \multicolumn{1}{c}{Issues (I)} & \multicolumn{1}{c}{Reactions (R)} &  \multicolumn{1}{c}{R/I} \\
\midrule
bug          & 289,386 \sbarf{289386}{862296} & 102,044 \sbarf{102044}{557696} & 0.35 \\
enhancement  & 213,475 \sbarf{213475}{862296} & 296,653 \sbarf{296653}{557696} & 1.39 \\
question     & 169,502 \sbarf{169502}{862296} & 63,621  \sbarf{63621}{557696}  & 0.38 \\
help         & 71,821  \sbarf{71821}{862296}  & 58,230  \sbarf{58230}{557696}  & 0.81 \\
duplicate    & 48,279  \sbarf{48279}{862296}  & 12,465  \sbarf{12465}{557696}  & 0.26 \\
stale        & 30,012  \sbarf{30012}{862296}  & 16,143  \sbarf{16143}{557696}  & 0.54 \\
invalid      & 26,081  \sbarf{26081}{862296}  & 2,024   \sbarf{2024}{557696}   & 0.08 \\
wontfix      & 13,740  \sbarf{13740}{862296}  & 6,516   \sbarf{6516}{557696}   & 0.47 \\ \midrule
Total        & \multicolumn{1}{c}{862,296}    & \multicolumn{1}{c}{557,696} & 0.53 \\
\bottomrule
\end{tabular}
\end{table}

\begin{figure}[!ht]
    \centering
    \includegraphics[page=1,width=\linewidth, trim={0 1em 2em 2em}, clip]{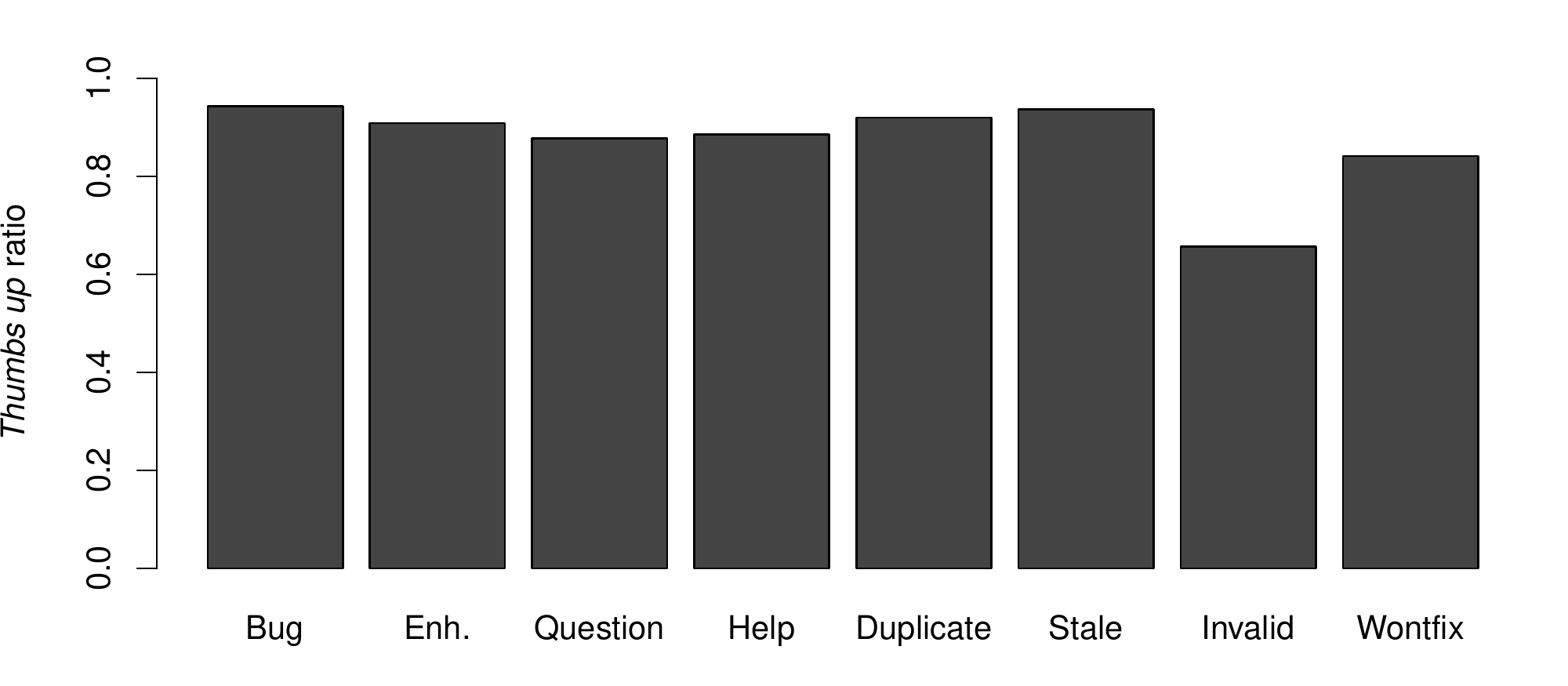}
    \caption{Usage of Thumbs up reaction by issue type}
    \label{fig:results:reactions-labels-thumbsup}
\end{figure}
\begin{figure}[!ht]
    \centering
    \includegraphics[page=2,width=\linewidth, trim={0 1em 2em 2em}, clip]{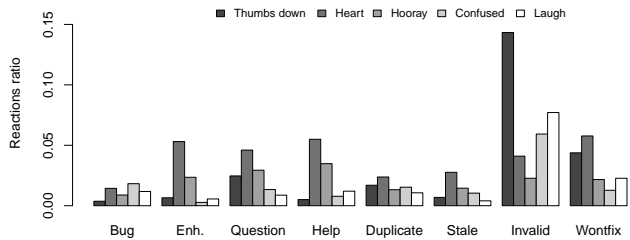}
    \caption{Usage of reactions by issue type}
    \label{fig:results:reactions-labels}
\end{figure}

Figure \ref{fig:results:reactions-labels-thumbsup} shows the percentage of \textit{Thumbs up} (the most common reaction in our dataset) per label, considering only labels with reactions. Excluding \textit{invalid}, almost all labels have at least 80\% of \textit{Thumbs up}, reaching the ratio of 94\% for \textit{bugs}. 
Finally, Figure \ref{fig:results:reactions-labels} compares the usage of the remaining reactions among the selected labels. \textit{Heart} presents a higher usage on issues broadly related to improvements, such as \textit{question} (4.6\%), \textit{help} (5.4\%), and \textit{enhancement} (5.3\%). \textit{Invalid} issues have a high number of \textit{Thumbs down} (14.3\%), \textit{Laugh} (7.7\%), and \textit{Confused} (5.9\%) reactions, which therefore are used by developers to express their level of disagreement with the issue content. Finally, \textit{wontfix} issues have higher levels of \textit{Thumbs down} (4.3\%) and \textit{Heart} (5.8\%), which reveals a conflicting sentiment regarding this issue type. 

\mybox{Summary}{Developers react more often to \textit{enhancements}. When we consider only \textit{bugs} and \textit{enhancements} with reactions, \textit{enhancements} have 1.39 reactions (on average), against 0.35 reactions in the case of \textit{bugs}. Moreover, there is a correlation between issue and reaction type. For example, \textit{Thumbs down}, \textit{Laugh} and \textit{Confused} are more common in invalid issues.}

\vspace{1em}\begin{tcolorbox}[halign=flush center, sharp corners,top=1pt,bottom=1pt]
Theme 2: Impact of Reactions
\end{tcolorbox}

\vspace{0.5em}\noindent\textit{RQ \#6: Do issues with more reactions get resolved faster?}\vspace{0.5em}

In this research question, we analyze the relation between the time to close a given issue and its number of reactions.
As discussed in RQ \#5, on GitHub issues can represent different purposes (e.g., bug and enhancements). 
First, we restrict our analysis to \textit{bugs} and \textit{enhancements} to evaluate how reactions impact the close time of such issues.
Also, we restrict this analysis to repositories with at least 102 issues (which is the median number of issues per repository in our dataset).
Moreover, since the complexity of issues may vary among repositories, we compute the time as $\mathit{T} = \mathit{T}\textsubscript{I} - \mathit{T}\textsubscript{R(I)}$, i.e., the difference (in days) between the time required to close a given issue ($\mathit{T}\textsubscript{I}$) and the median time to close the other issues in the same repository ($\mathit{T}\textsubscript{R(I)}$).

\begin{figure}[!ht]
    \centering
    \includegraphics[width=.975\linewidth, trim={0 1em 1em 2em}, clip,page=2]{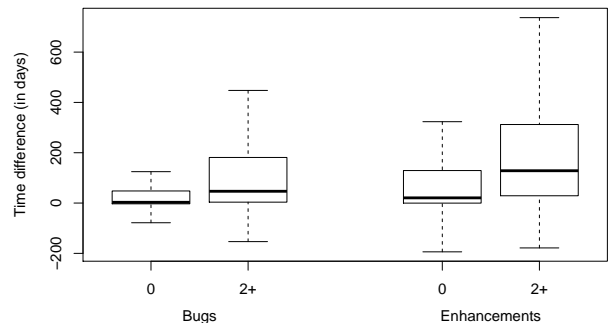}
    \caption{Time to close \textit{bugs} and \textit{enhancements} (with zero or with at least two reactions). Outliers are omitted}
    \label{fig:results:time}
\end{figure}

Figure~\ref{fig:results:time} presents the distribution of $\mathit{T}$ for \textit{bugs} and \textit{enhancements}, for issues without reactions and for issues with two or more reactions (which is the median number of reactions in our dataset). The results suggest that issues with reactions take more time to be closed than the ones without reactions. For example, \textit{bugs} without reactions take 2.97 more days to be closed, compared with the repository median time, whereas the ones with at least two reactions take 46.8 more days. For \textit{enhancements}, the median values are 20.64 and 128.45 days, respectively.
By applying the Mann-Whitney test, we concluded that the distributions are different (\emph{p-value} $< 0.001$). By computing Cohen's $d$, we found a \textit{medium} effect size for \textit{bugs} and \textit{enhancements} ($d = 0.511$ and $0.572$, respectively).

\begin{figure}[!ht]
    \centering
    \includegraphics[width=.975\linewidth, trim={0 1em 1em 2em}, clip,page=2]{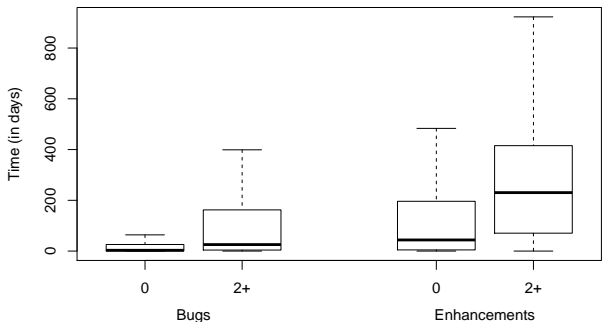}
    \caption{Time to close \textit{bugs} and \textit{enhancements} in \textsc{Microsoft/vscode} (outliers are omitted)}
    \label{fig:results:time-microsoft-vscode}
\end{figure}

In fact, Figure~\ref{fig:results:time-microsoft-vscode} presents the distribution of the time taken to close \textit{bug} and \textit{enhancement} issues in \textsc{Microsoft/vscode}, which is the repository with the highest number of reactions in our dataset. In this specific repository, we also observe that \textit{bugs} and \textit{enhancements} with at least two reactions take more time to be closed compared to the ones without reactions.
On median, \textit{bugs} without reactions are closed in 3.1 days whereas those ones with at least two reactions are closed in 25.5 days. Regarding \textit{enhancements}, the median times are 43.8 and 230.3 days, respectively.

\vspace{0.5em}\mybox{Summary}{\textit{Bugs} and \textit{enhancements} with reactions take more time to be closed. We hypothesize that they are more complex and require more time from developers to be resolved.}

\vspace{0.5em}\noindent\textit{RQ \#7: Do issues with more reactions have more discussion?}\vspace{0.5em}

In the previous research question, we analyzed the relationship between reactions and the time to close issues.
Here, we analyze whether the number of reactions is associated to longer discussions (i.e., more comments). 
To this purpose, we first compute the number of comments in \textit{bugs} and \textit{enhancements}. Then, we compare the results for issues without reactions and with at least two reactions.

\begin{figure}[!ht]
    \centering
    \includegraphics[width=.975\linewidth, trim={0 1em 1em 2em}, clip,page=4]{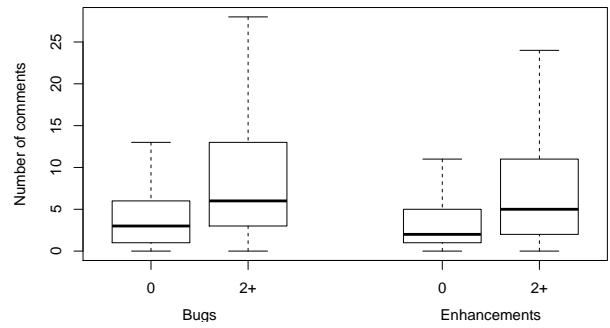}
    \caption{Number of comments in \textit{bugs} and \textit{enhancements} (without reactions vs with at least two reactions). Outliers are omitted}
    \label{fig:results:comments}
\end{figure}

Figure~\ref{fig:results:comments} presents these distributions. In all cases, issues with reactions have more comments than those without reactions. For example, \textit{bugs} without reactions have 3 comments against 6 comments (median) in the case of \textit{bugs} with at least two reactions. Besides that, \textit{enhancements} without reactions have 2 comments (median), whereas \textit{enhancements} with reactions have at least 5 comments.
By applying a Mann-Whitney test, we detected that the distributions are different (\emph{p-value} $<$ 0.001).
We also found a \textit{large} effect size between the distributions of \textit{bugs} (Cohen's $d = 0.943$). For \textit{enhancements}, we found a \textit{medium} effect size (Cohen's $d = 0.626$).

\begin{figure}[!ht]
    \centering
    \includegraphics[width=.975\linewidth, trim={0 1em 1em 2em}, clip,page=4]{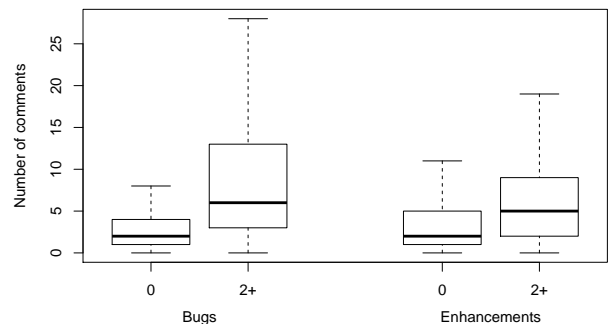}
    \caption{Number of comments in \textit{bugs} and \textit{enhancements}, in \textsc{Microsoft/vscode} (outliers are omitted)}
    \label{fig:results:comments-microsoft-vscode}
\end{figure}

Figure~\ref{fig:results:comments-microsoft-vscode} shows the distribution of the number of comments in \textit{bugs} and \textit{enhancements} of a specific repository, \textsc{Microsoft/vscode}.
In this project, we confirm that issues with reactions have more comments. The median number of comments in \textit{bugs} is 2 and 6, respectively. For \textit{enhancements}, the median values are respectively 2 and 5 comments, respectively.

\vspace{0.5em}\mybox{Summary}{\textit{Bugs} and \textit{enhancements} with reactions have longer discussion, which essentially confirms the result of RQ \#6, i.e., reactions tend to appear in more complex issues, which demand more discussion.}

\vspace{0.5em}\noindent\textit{RQ \#8: Do negative reactions inhibit further participation?}\medskip

In the literature, several authors argue that negative feedback is a factor that inhibits users participation in social platforms~\cite{lampe2005follow, Katmada2016,Aniche2018}.
Here, we aim to investigate whether negative reactions (i.e., \textit{Thumbs down}) inhibits further participation through issue reports and comments.
Specifically, we compute the number of interactions (i.e., issues and comments) performed by developers \textit{before} and \textit{after} receiving negative reactions. 
Also, we restrict our analysis to issues and comments that have more \textit{Thumbs down} than the sum of other reactions. When a developer has more than one issue or comment with the required number of \textit{Thumbs down}, we selected the most negative one. Here, we analyze the developers' participation in a project, i.e., the interactions are computed in the same repository with the negative reactions.

\begin{figure}[!ht]
    \centering
    \includegraphics[width=.8\linewidth, clip]{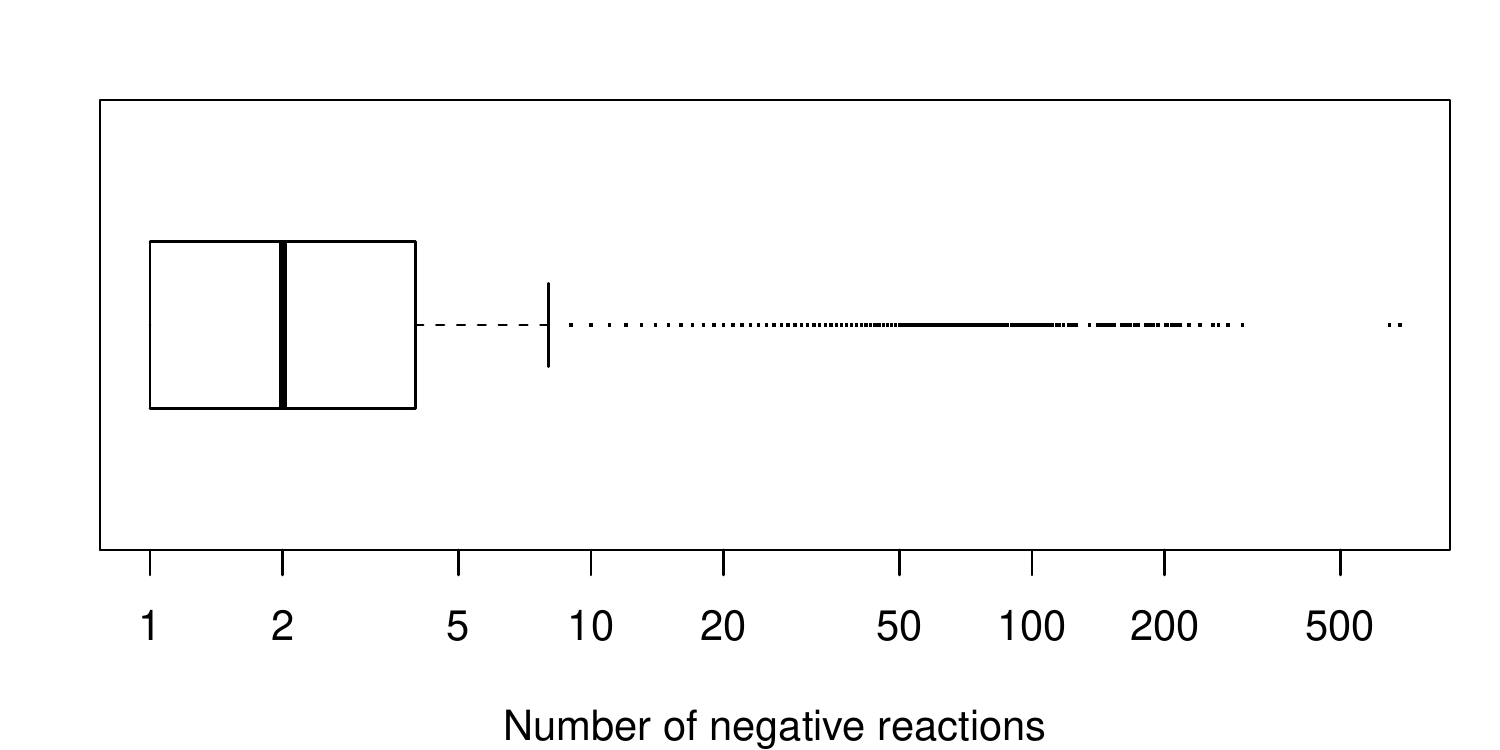}
    \caption{Distribution of negative reactions in issues and comments}
    \label{fig:results:negative-issues-distribution}
\end{figure}

First, Figure~\ref{fig:results:negative-issues-distribution} presents the distribution of the number of \textit{Thumbs down} reactions received by developers. The distribution shows that most developers receive a low number of negative feedback. However, other developers receive a considerable number of \textit{Thumbs down}. For example, we have 22,431 developers with at least one issue or comment with more \textit{Thumbs down} than other reactions. Therefore, in this research question, we focus on the top-10\% issues and comments, which consists of 2,243 issues and comments with values that range from 11 to 683 \textit{Thumbs down} reactions.

\begin{figure}[!ht]
    \centering
    \includegraphics[width=.8\linewidth, clip]{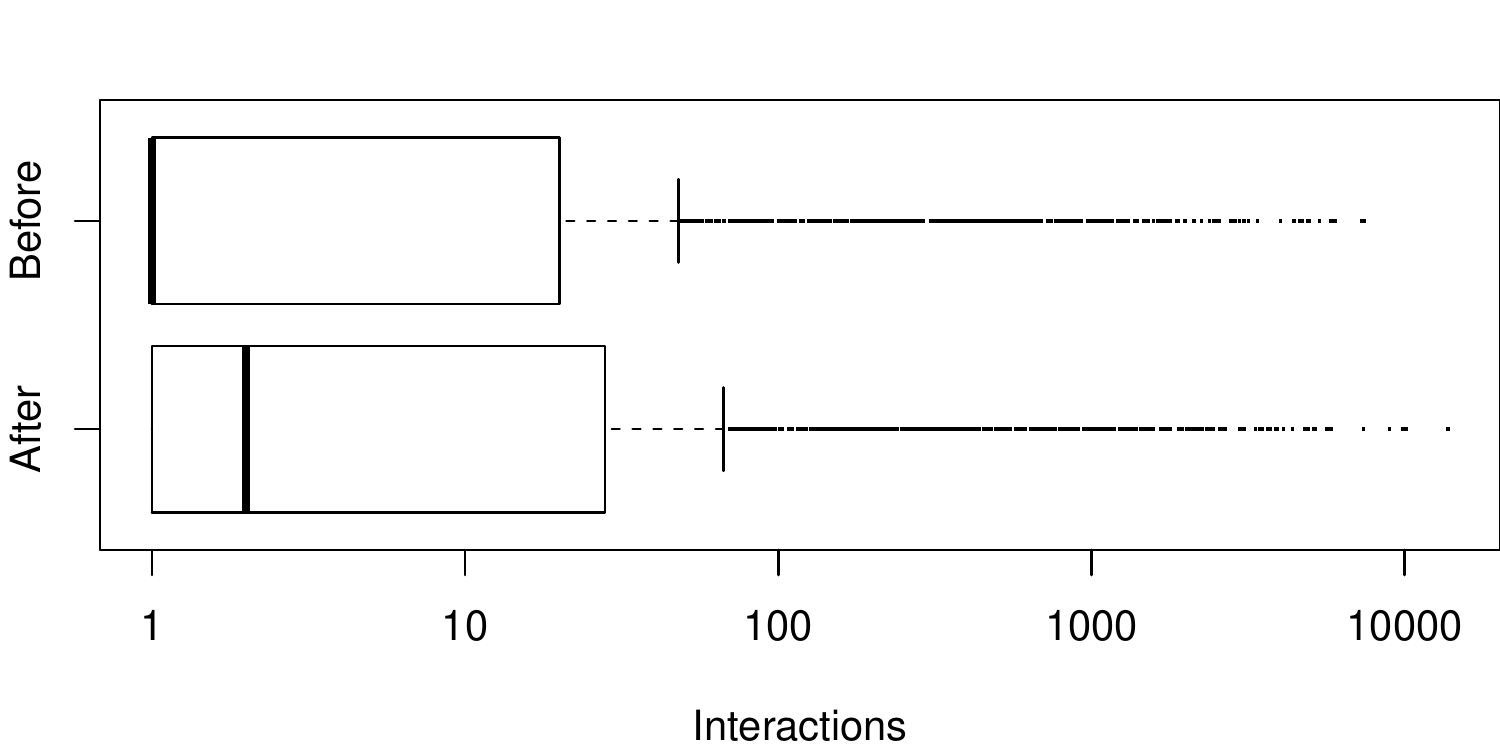}
    \caption{Number of issues and comments made by developers before and after receiving a high number of \textit{Thumbs down}}
    \label{fig:results:negative-results}
\end{figure}

In Figure~\ref{fig:results:negative-results}, we present the distribution of the number of interactions (i.e., issue reports or issue comments) made \textit{before} and \textit{after} negative reactions. We observe that the values are very close. For example, the median values are 1 and 2 interactions before and after the negative issue reports or comments.
By applying the Mann-Whitney test, we detected that these distributions are different (p-value $<$ 0.001), but with a \textit{very} small effect size (Cohen's d = 0.136). 

\vspace{0.5em}\mybox{Summary}{Differently from other social platforms, negative reactions do not inhibit GitHub developers participation in further issues of the same repository.}

\vspace{0.5em}\noindent\textit{RQ \#9: Do reactions reduce the noise in issue conversations?}\vspace{0.5em}

When the reactions feature was introduced, GitHub argued that one of its purposes was to reduce the noise caused by emojis in issue responses. 
Indeed, before the introduction of reactions, developers often added comments with a single \textit{Thumbs up}. 
In this research question, we investigate whether the introduction of reactions reduced the use of emojis (e.g., \thumbsup,\thumbsdown, etc.) in issue conversations.
To this purpose, we analyzed the issues created after the introduction of reactions and that have emojis in their body. Specifically, we searched for comments containing emoji markups in the format \texttt{:<id>:}, where \texttt{<id>} refers to the emoji's code. For example, \texttt{:thumbsup:} and \texttt{:+1:} refer to the emoji~\thumbsup. 

\begin{figure}[!ht]
    \centering
    \includegraphics[width=.925\linewidth, trim={0 0 0 2.5em}, clip, page=1]{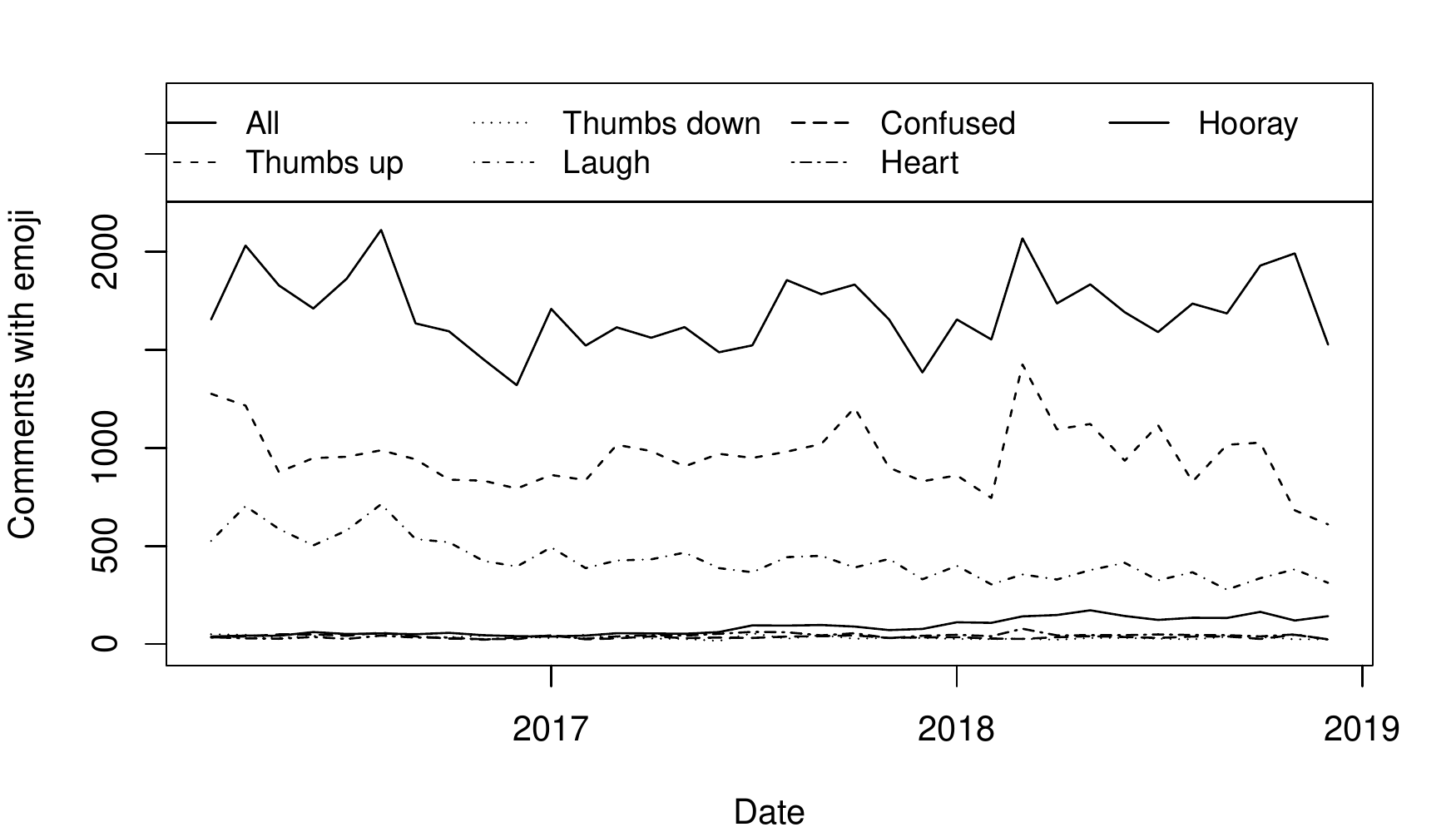}
    \caption{Usage of emojis in the body of comments after reactions introduction}
    \label{fig:results:emoji}
\end{figure}

Figure~\ref{fig:results:emoji} presents the monthly usage of emojis in the body of comments, considering all available emojis\footnote{\url{https://gist.github.com/rxaviers/7360908}, verified on 01/20/2019.} and only the emojis representing GitHub reactions (i.e., \thumbsup,\thumbsdown,\hooray,\heart,\confused, and \laugh). The results suggest that the usage of emojis in comments remains the same. In addition, similar results are observed for emojis associated to reactions. The only exception is the emoji \texttt{:tada:} (\hooray), whose usage increased considerably in the last years. Thus, we can conclude that GitHub reactions did not reduce the usage of emojis in the body of comments.

\begin{figure}[!ht]
    \centering
    \includegraphics[width=\linewidth, trim={0 0 0 2.5em}, clip, page=2]{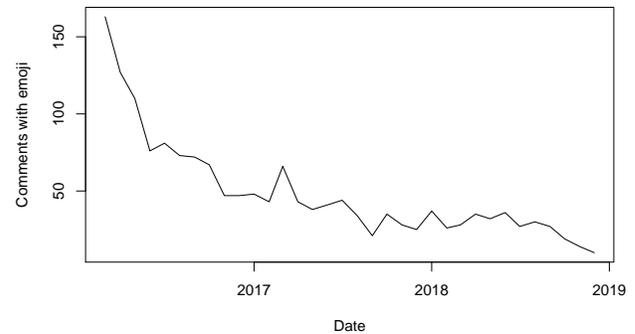}
    \caption{Usage of only \textit{Thumbs up} in comments after reactions introduction}
    \label{fig:results:emoji-thumbs-up}
\end{figure}

Finally, we analyzed the comments containing only a single emoji in their body. Specifically, we found that only \textit{Thumbs up} (\thumbsup) and \textit{Heart} (\heart) are used in this manner. However, \textit{Heart} is used in only 18 comments, whereas \textit{Thumbs up} is used in 1,650 comments. Figure~\ref{fig:results:emoji-thumbs-up} shows the monthly usage of comments containing a single \textit{Thumbs up}. We can conclude that this practice  is in sharp decrease on GitHub. 

\mybox{Summary}{GitHub developers are still using emojis in their comments; however they are no longer commenting with a single \thumbsup.}
\section{Key Findings and Implications}
\label{sec:key-finding-and-implications}

We found that most issue reports (87\%), comments (89\%), and threads (71.6\%) do not have reactions (RQ \#1). However, it is important to acknowledge that the usage of reactions is gaining momentum; it increased  32.5\% in 2018, compared to 2017 (RQ \#3). We also reported that {\em Thumbs up} is by far the most popular reaction (78.7\%, RQ \#2).
 
\vspace{0.75em}\noindent{\em Practical Recommendation \#1:} We reported interesting use cases of reactions, which might be of the interest of users and owners of GitHub projects. Mostly, we found that reactions can be used  (1) to express the importance of fixing a bug or implementing a new feature; (2) to implement surveys with users (e.g., a logo context); (3) to collect user's feedback about a new release candidate, before it is deployed; (4) to indicate that an issue is invalid. In all these scenarios, reactions are instruments to collect user feedback. But in scenarios (1), (3), and (4), the feedback is spontaneous, i.e., the users are not asked to react. By contrast, in scenario (2), the manifestation is explicitly requested by project managers.
\vspace{0.75em}
 
We also found that reactions and stars do not have a strong correlation (RQ \#4). Interestingly, we found that users react four times more to enhancements than to bugs (RQ \#5). We claim that this finding reveals the pressure suffered by GitHub maintainers to continuously evolve their systems, by providing new features. Furthermore, GitHub users tend to react to more complex issues and enhancements, i.e., the ones that take more time to be handled. For example, enhancements with at least two reactions take 108 more days to be implemented, regarding the median implementation time of other enhancements---without reactions---in a repository.
 
\vspace{0.75em}\noindent{\em Practical Recommendation \#2:} GitHub users should not expect that by reacting to an issue they will get it handled in a few days. Usually, simple issues are handled quickly, even when they do not have reactions. By contrast, it is natural that complex issues require more time to be handled, even when they have many {\em Thumbs up}. This is exactly the case of the most  reacted issue in our dataset, which refers to a new feature which is yet not supported by Microsoft Visual Studio. In other words, reactions should not be viewed by users as an instrument capable of accelerating the implementation of bug fixes or new features. They are not an instrument of pressure; but an instrument to provide feedback.
\vspace{0.75em}

We also found that issues with reactions have longer discussions (RQ \#7) and that negative reactions (\textit{Thumbs down}) do not inhibit developers to comment again (RQ \#8).

\vspace{0.75em}\noindent{\em Practical Recommendation \#3:} GitHub developers should not view negative reactions as an invitation to abandon projects. Most developers continue to participate in a project, even after receiving a massive number of {\em Thumbs down}.
\vspace{0.75em}

Finally, we showed that reactions are not contributing to reduce emojis in comments, including the ones used by the existing GitHub reactions (RQ \#9). However, the number of comments with a single \textit{Thumbs Up} is in sharp decline.

\vspace{0.75em}\noindent{\em Practical Recommendation \#4:} GitHub should consider supporting other emojis; for example \textit{Clapping hands} (\clap), whose usage is facing a constant increment. Another possible feature is to allow developers to highlight particular sentences of an issue or comment, and then apply a reaction only in the highlighted text. These fine-grained reactions can contribute to reduce the noise caused by emojis in the body of comments (which is not declining, as  we showed in RQ \#9). Finally, traditional bug tracking systems may also benefit from reactions in their comment systems.
\section{Threats to Validity}
\label{sec:threats-to-validity}

\noindent\textit{Construct Validity: }GitHub has millions of repositories. We built our dataset by collecting the top-5,000 repositories by number of stars, which represents a small fraction in comparison to the GitHub's universe.
However, most GitHub repositories are forks, have very low activity, or have a small community~\cite{Kalliamvakou2014, Kalliamvakou2015, cosentino2017}. Despite the repository selection by number of stars, our dataset includes projects from a vast range of programming languages and domains.

\vspace{0.5em}\noindent\textit{Internal Validity: } An internal threat relates to the issue types used in  RQ \#5. In this research question, we relied on the labels provided by the projects maintainers to characterize the issue type. Moreover, we also performed an extra step to group labels with the same purpose. To mitigate this threat, we restricted our analysis to a small group of labels, which is also recommended by default on GitHub. Also, the labels groups were manually verified by the first and second authors of this paper to guarantee that those in the group refer to the same label.

\vspace{0.5em}\noindent\textit{External Validity: }The data analyzed in this work is from the GitHub platform. Since the reactions feature is exclusive of this platform, the results may not generalize to other social coding platforms (e.g. GitLab and BitBucket).
Finally, during the development of this research, GitHub released two new reactions: Rocket(\rocket) and Eyes(\eyes). The inclusion of new reactions may influence the usage of  reactions in the future. However, we claim these new reactions do not have an immediate impact; and they should only be studied after gaining some adoption~\cite{rogers2010diffusion}.
\section{Related Work}
\label{sec:related}

In the literature, several studies explore emotions on software development. Murgia \textit{et al.} \cite{murgia2014} analyzed whether issue reports carry emotional information. Their study confirms that issue reports do express emotions towards design choices, maintenance activity or colleagues. Jongeling \textit{et al.}~\cite{jongeling2015} investigated the agreement between sentiment analysis tools and the sentiment recognized by humans. Their results suggest a sentiment analysis tool specially targeting the software engineering domain is needed. 
Brants \textit{et al.}~\cite{brants2019} conducted a survey in order to determine whether people interpret emojis (specific to emotional states) in a consistent manner. They conclude that there are emojis that are interpreted differently depending on age or gender.

Other studies explore Github projects. Borges \textit{et al.} \cite{borges2016} investigated the characteristics of popular GitHub repositories and found four major popularity growth patterns. 
Recently, Coelho \textit{et al.} \cite{jailton2018} proposed an approach to identify unmaintained projects in GitHub. Their approach apply machine learning techniques and rely on metrics that measure the maintained level to identify and alert users. Machine learning is also used in GitHub to classify user's sentiments in the comments of issues \cite{Werder2018, Ding2018} and to identify software vulnerabilities \cite{Harer2018}. 
User behavior is also analyzed in other studies. Jiang \textit{et al.} \cite{Jiang2019} presented an analysis of unfollowing events between GitHub users, identifying potential factors that can motivate this behaviour.
Bissyand\'{e} \textit{et al.} \cite{Bissyande2013} performed an empirical study about adoption of issues tracker in software projects. The authors investigated the involvement of users and developers when reporting issues and their impact on project success. 

Moreover, several studies investigate the impact of reactions--- such as \textit{upvotes}, \textit{favorite}, and \textit{like}---in other social networks. In a recent study, Aniche \textit{et al.} \cite{Aniche2018} investigated two popular news aggregators sites (HackerNews and Reddit). The authors showed that developers view \textit{upvotes} and \textit{comments} as measurement of post's quality and relevance. Likewise, Sharma \textit{et al.} \cite{Sharma2017} proposed a recommendation system to support developers learning through URLs posted on Twitter. Their system relies on users actions, as \textit{favorite} and \textit{retweet}, to identify and rank relevant content.

\section{Conclusion}
\label{sec:conclusion}

In this paper, we investigated how developers react to issue reports and their comments on GitHub.
For this purpose, we collected and analyzed the data of 4,849 public GitHub projects.
We provided answers to nine research questions, referring to two  themes: (i) the usage of reactions by GitHub developers, and (ii) the impact of reactions on software development practices.
Our key findings are as follows.

\begin{itemize}
    \item Most issues (87\%) and comments (89\%) do not have reactions. For whole threads (issues + comments), we found that 28.4\% have at least one reaction. Moreover, \textit{Thumbs up} is by far the most common reaction (78.7\%).
    
    \item Reactions are gaining momentum; in 2018, developers used 32.5\% more reactions than in 2017. 
    
    \item We show that popular projects do not necessarily attract more reactions. We also show that developers react more often to \textit{enhancements} and there is a correlation between issue types and reactions.
    
    \item \textit{Bugs} and \textit{enhancements} with reactions take more time to be closed and have longer discussion.
    
    \item On the contrary of other social platforms, negative reactions do not inhibit GitHub developers participation in further issues of the same repository.
    
    \item Finally, GitHub reactions are contributing to reduce the number of comments containing a single \textit{Thumbs up} emoji. However, developers are still using emojis in their comments.
\end{itemize}

As future work, we plan to investigate the impact of the profile of the analyzed projects on the usage of reactions. We also plan to conduct studies with GitHub developers to understand their perceptions on the reactions features. First, we intend to conduct a survey to discover what are the factors they consider before reacting to issues and to comments.
We also plan to shed light on the importance of reactions for project maintainers through an opened survey with developers from popular projects.
Finally, we plan to extend our work to consider reactions on pull requests.

The dataset with the repositories, issues, comments, and reactions used in this paper is publicly available at: \url{https://doi.org/10.5281/zenodo.2558596}.

\section*{Acknowledgments}
\noindent This research is supported by grants from FAPEMIG, CAPES, and CNPq.

\bibliographystyle{acm}
\bibliography{references}

\end{document}